\documentclass{article}

\usepackage{arxiv}
\usepackage[utf8]{inputenc} % allow utf-8 input
\usepackage[T1]{fontenc}    % use 8-bit T1 fonts
\usepackage[colorlinks,citecolor=blue,urlcolor=red]{hyperref}
\usepackage{url}            % simple URL typesetting
\usepackage{booktabs}       % professional-quality tables
\usepackage{amsfonts}       % blackboard math symbols
\usepackage{nicefrac}       % compact symbols for 1/2, etc.
\usepackage{microtype}      % microtypography
\usepackage{graphicx}
\usepackage{natbib}
\usepackage{doi}
\usepackage{amsmath}
\usepackage[figuresright]{rotating}
\usepackage{color}
\usepackage{comment}

\title{Assessing Delayed Treatment Benefits of Immunotherapy Using Long-Term Average Hazard: A Novel Test/Estimation Approach}

\author{
    {Miki~Horiguchi}\\
    Department of Medical Oncology\\
	Dana Farber Cancer Institute\\
	Boston, MA 02215, USA \\
	\And
    {Lu~Tian}\\
	Department of Biomedical Data Science\\
    Stanford University\\ 
    Stanford, CA 94305, USA \\
    \And
    {Kenneth~L.~Kehl}\\
    Department of Medical Oncology\\
	Dana Farber Cancer Institute\\
	Boston, MA 02215, USA \\
    \And
    {Hajime~Uno}\thanks{Email: huno@ds.dfci.harvard.edu} \\
	Department of Data Science\\
    Department of Medical Oncology\\
	Dana Farber Cancer Institute\\
	Boston, MA 02215, USA \\
}

\date{}

\renewcommand{\headeright}{}
\renewcommand{\undertitle}{}
\renewcommand{\shorttitle}{Assessing Long-Term Treatment Effect}

\hypersetup{
pdftitle={AH long term},
pdfsubject={statME},
pdfauthor={Horiguchi, Tian, Kehl, Uno},
pdfkeywords={AH, cGFIR},
}

\begin{document}

\maketitle

\baselineskip=15pt
\begin{abstract}
Delayed treatment effects on time-to-event outcomes have often been observed in randomized controlled studies of cancer immunotherapies. In the case of delayed onset of treatment effect, the conventional test/estimation approach using the log-rank test for between-group comparison and Cox's hazard ratio to estimate the magnitude of treatment effect is not optimal, because the log-rank test is not the most powerful option, and the interpretation of the resulting hazard ratio is not obvious. Recently, alternative test/estimation approaches were proposed to address both the power issue and the interpretation problems of the conventional approach.  One is a test/estimation approach based on long-term restricted mean survival time, and the other approach is based on average hazard with survival weight.  This paper integrates these two ideas and proposes a novel test/estimation approach based on long-term average hazard (LT-AH) with survival weight. Numerical studies reveal specific scenarios where the proposed LT-AH method provides a higher power than the two alternative approaches. The proposed approach has test/estimation coherency and can provide robust estimates of the magnitude of treatment effect not dependent on study-specific censoring time distribution. Also, the proposed LT-AH approach can summarize the magnitude of the treatment effect in both absolute difference and relative terms using ``hazard'' (i.e., difference in LT-AH and ratio of LT-AH), meeting guideline recommendations and practical needs. This proposed approach can be a useful alternative to the traditional hazard-based test/estimation approach when delayed onset of survival benefit is expected.
\end{abstract}


\clearpage
%======================
\section{Introduction}
%======================

%------------------------------------------------------ 
% 1. Introduction of non-PH (immunotherapy trials)
%------------------------------------------------------ 
In recent years, there have been many immunotherapy clinical trials in cancer clinical research.
According to {\it ClinicalTials.gov}, as of October 1, 2023, 86 phase III clinical trials were actively recruiting patients to evaluate checkpoint inhibitor immunotherapies.
Generally, in late-stage cancer clinical trials, the primary outcomes for evaluating the treatment effect of investigational therapies are time-to-event outcomes such as overall survival and progression-free survival (PFS), and the conventional analytical approach uses the log-rank test for statistical comparison and Cox's hazard ratio to estimate the magnitude of the treatment effect.
On the other hand, various authors have argued that this conventional log-rank/hazard-ratio test/estimation approach is not optimal for immunotherapy trials because many empirical studies have shown that the proportional hazards assumption does not hold in these trials \citep{Chen2013-dy, Xu:2017dz, Alexander2018-vt, Huang2018}. 
The pattern of difference that is often observed in immunotherapy trials is the so-called {\it `delayed difference.'} 
Figure 1(A) illustrates an example of the delayed difference pattern. 
It shows the Kaplan-Meier curves of PFS data from a randomized controlled trial that compared nivolumab plus ipilimumab to sunitinib in patients with advanced renal cell carcinoma \citep{Motzer2018-ie}. 
In this trial, the two PFS curves were almost identical from time 0 to 7 months, but a benefit of immunotherapy appeared after 7 months. 
Of course, if the sample size of the study is not so large, it would be possible for two estimated survival curves to have a delayed difference pattern even when the sample is from a proportional hazards data generation process.  
However, given the empirical evidence from previous immunotherapy trials, it would not be best practice to use the conventional log-rank/hazard-ratio test/estimation approach, which relies on the proportional hazards assumption. 

%%%%%%%%%%%%%%%%%%%%%%%%%%%%%%%%%%%%
% Figure 1: Example
%%%%%%%%%%%%%%%%%%%%%%%%%%%%%%%%%%%%
%\clearpage
%\newpage
\begin{figure}[h]
\label{figure1}
\centering
\includegraphics[scale=0.5 ]{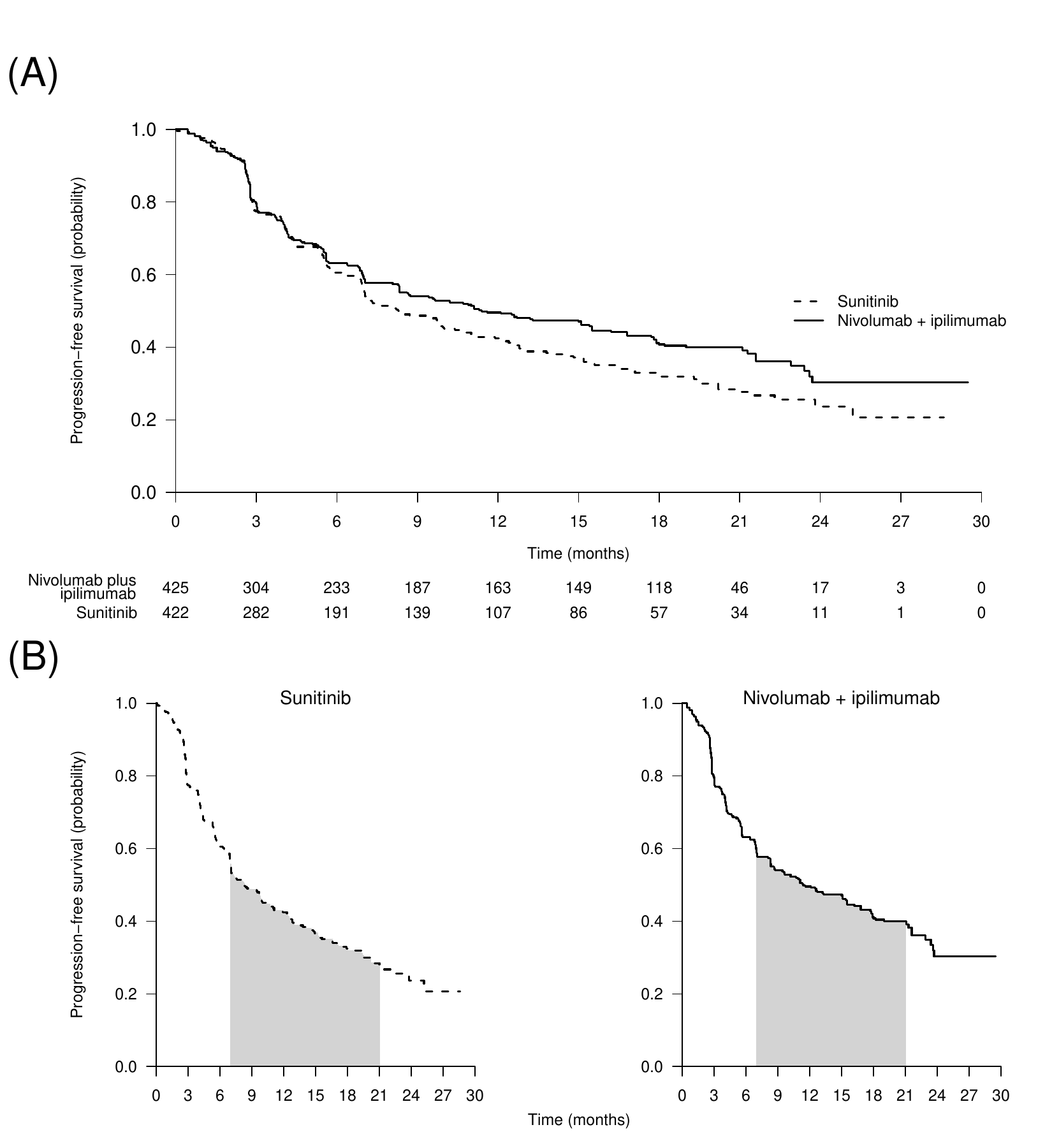}
\caption{Estimated Kaplan-Meier curves of progression-free survival time (A) and the area under the curves on the time range [7, 21] (months) (B) with the data reconstructed from the publication of the CheckMate 214 study.}
\end{figure}

%------------------------------------------------------ 
% 2. The log-rank/hazard ratio is not appropriate to use.
%------------------------------------------------------ 
Here, we briefly describe some more general considerations with respect to the conventional log-rank/hazard-ratio test/estimation approach. First, this conventional approach exhibits test/estimation coherency \citep{Uno2020-gx}. 
The statistical significance derived from the test result will not contradict whether the confidence interval for the hazard ratio covers the null value or not. 
Since the qualitative conclusion derived from the test and the quantitative summary derived from the interval estimation are coherent, this conventional approach will not generate confusing results for decision-making.  

  %--------------------
  % 2.1 Testing
  %--------------------
Regarding the testing part of the conventional test/estimation approach, the log-rank test and the partial likelihood-based tests (such as the score test and the Wald test) for the hazard ratio are asymptotically valid tests regardless of the adequacy of the proportional hazards assumption. Also, if the proportional hazards assumption holds, these tests are asymptotically the most powerful tests. On the other hand, when the proportional hazards assumption does not hold, they are not optimal \citep{FH1991}. For the delayed difference pattern, for example, other tests will offer higher power than these. 

  %--------------------
  % 2.2 Estimation
  %--------------------
Regarding the estimation part of the conventional test/estimation approach, Cox's method provides an efficient and elegant way to summarize the magnitude of the treatment effect as a hazard ratio. It does not require a distribution assumption for survival time in each group. Also, it does not require estimation of the group-specific absolute hazard to calculate the hazard ratio; it can directly estimate the hazard ratio by imposing the proportional hazards assumption. 
However, this approach also has several limitations. 
One notable limitation is that, when the proportional hazards assumption does not hold, Cox's hazard ratio depends on the underlying study-specific censoring time distribution, and therefore the interpretation of the resulting hazard ratio is not obvious \citep{Kalbfleisch1981-xo,Lin:1989vg,Uno:2014ii,Horiguchi2019-xc}.
Another notable limitation is the lack of group-specific absolute hazards. Although this may be an advantage from a statistical point of view, it can be a disadvantage from a practical point of view. The same hazard ratio will have a different clinical implication if the absolute hazard in the control group is different. Therefore, the group-specific absolute hazards will be necessary information for clinicians to assess if the resulting hazard ratio indicates a clinically significant magnitude of the treatment effect. 
In fact, some guidelines recommend presenting the magnitude of a treatment effect in both absolute difference and relative terms for better clinical interpretation. 
For example, in the CONSORT 2010 guideline, Subitem 17b says {\it ``When the primary outcome is binary, both the relative effect (risk ratio (relative risk) or odds ratio) and the absolute effect (risk difference) should be reported (with confidence intervals), as neither the relative measure nor the absolute measure alone gives a complete picture of the effect and its implications.''} \citep{Cobos-Carbo2011-wz} 
While Subitem 17b is specific for binary outcomes, this recommendation also applies to time-to-event outcomes. Similarly, the General Statistical Guidance provided by the Annals of Internal Medicine provides comparable guidance for time-to-event outcomes: {\it ``Presenting estimates of effect in both absolute and relative terms increases the likelihood that results will be correctly interpreted.''} (\citeauthor{annals}, 2024) 
Unfortunately, except for very specific cases, it is difficult to transform the hazard ratio from Cox's proportional hazards model into the absolute difference in hazard. 

%------------------------------------------------------ 
% 3. Review of the alternative approaches 
%------------------------------------------------------ 

  %------------------------------------------------------ 
  % 3.1 WLR class
  %------------------------------------------------------ 
There are many alternative approaches to address these limitations of the conventional approach. With respect to power loss under non-proportional hazards scenarios in the testing part of the conventional approach, one popular alternative is to use a test from a class of weighted log-rank tests. For example, for the delayed separation pattern shown in Figure 1(A), $G^{0, 1}$ test in the $G^{\rho, \gamma}$ class \citep{FH1991} will be more powerful than the ordinal log-rank test (i.e., $G^{0, 0}$).  The piecewise weighted log-rank test proposed by \cite{Xu:2017dz} is another alternative when a delayed separation pattern is expected. This test will give zero weight to the early time window $(0, \eta)$ where two survival curves seem identical. 

Although a delayed separation pattern has been observed in most immunotherapy trials, this may not always be the case. 
To address the possible misidentification of the pattern of difference between two survival curves, robust (or versatile) tests using several weighted log-rank test statistics have been proposed \citep{Tarone:1977ux,FH1991} to capture various patterns of difference.  For example, a cross-pharma working group \citep{Roychoudhury2021-bv} for clinical trials with non-proportional hazards recommended the so-called {\it Max-combo} test, which uses the maximum of the test statistics of $G^{0,0},$ $G^{1,0},$ $G^{0,1}$ and $G^{1,1}$ as the test statistic.

These alternatives using weighted log-rank tests will provide higher power than the conventional log-rank test for the trials where the appearance of a delayed treatment benefit is expected. However, one of the drawbacks of these is the estimation of the magnitude of the treatment effect. The treatment effect summary measure that corresponds to a weighted log-rank test will be an hazard ratio-type measure \citep{Leon2020-ps} and thus has the same issues as Cox's hazard ratio as described.

  %------------------------------------------------------ 
  % 3.2 t-year, Median, RMST (non-parametric trio)
  %------------------------------------------------------ 
In addition to weighted log-rank approaches, a wide array of statistical methods has been developed to address non-proportional hazards, with a comprehensive review of these methods detailed in the work of \cite{Bardo2023-qa}.
In this paper, however, we focus on a specific class of alternative approaches that seek to overcome the limitation regarding not only non-proportional hazards but also the limitation regarding the lack of group-specific absolute hazards. 
Specifically, the class of alternative approaches we are interested in here uses summary measures of the event time distribution for each group. This allows the derivation of control measures that quantify between-group differences, encompassing both absolute and relative metrics \citep{Uno:2014ii,Chappell:2016dm}.
For example, the cumulative incidence probability at a specific time point or the restricted mean survival time (RMST) with a specific truncation time point \citep{Royston2011-qx,Uno:2014ii,Uno:2015ip,AHern:2016df,Chappell:2016dm,Peron2016-nk,Saad2017-mh} can be used as the summary measure of the event time distribution. 
Median survival time can also be used if it is estimable in both groups. Of these, the approach using RMST is gaining more attention currently. Randomized trials where the RMST-based approach was used for the analysis can be found in the literature \citep{Guimaraes2020-zf,Connolly2022-sx, Hammad2022-ap, Sanchis2023-bl}. However, the RMST-based approach also has a limitation. Unfortunately, the statistical comparison based on RMST provides lower power than the conventional log-rank test when detecting delayed treatment effects \citep{Tian2018-hc}.

%------------------------------------------------------ 
% 4. Modification of RMST for Delayed separation (two directions)
%------------------------------------------------------ 

There are two directions one could take in order to address the power issue of the RMST-based approach under delayed difference patterns. 
Let $S(t)$ be the survival function of the event time $T.$ 
The {\it standard RMST} integrates the survival function with respect to time from time 0 to a specific truncation time point $\tau,$ which can be denoted by
$\int_0^{\tau} S(u)du.$
One direction to improve the power to detect a delayed treatment effect is to modify the time range in the calculation of the RMST as follows,
$\int_{\eta}^{\tau} S(u)du,$
where $\eta$ will be a positive constant and near the time point where two survival curves start to separate. 
As shown in Figure 1(A), the two survival curves are almost identical for the first 6 to 7 months.
Because $\int_{0}^{\eta} S(u)du$ provides mainly noise rather than a signal for statistical comparison between two groups, using $\int_{\eta}^{\tau} S(u)du$ will improve power. 
In this paper, to distinguish this from the standard RMST, we call this {\it Long-Term RMST.} 
A non-parametric inference procedure for this measure was given by \cite{Zhao2011-zg}, and an application to an immunotherapy study is found in \cite{Horiguchi:2018dk}. Recently, \cite{Paukner2021-vm} also introduced this approach, calling it {\it Window RMST.} 

Another direction to address the power issue of the standard RMST approach in detecting the delay effect of treatment is to modify the summary measure. 
Recently, \cite{Uno2023-sm} proposed a new summary measure of the event time distribution called ``{\it average hazard with survival weight}'' and nonparametric inference procedure for the difference and ratio of this metric.
The average hazard with survival weight is defined as  
\begin{equation} 
\frac{\int_{0}^{\tau} h(u)S(u)du}{\int_{0}^{\tau} S(u)du} = \frac{1-S(\tau)}{\int_{0}^{\tau} S(u)du} = \frac{E\{I(T\le \tau)\}}{E\{ T \wedge \tau \}},
\label{AH}
\end{equation}
where $h(u)$ is the hazard function for the event time $T,$ 
$I(A)$ is an indicator function for the event $A,$
and $x \wedge y $ denotes $\min(x,y).$
Looking at the last term in the equation (\ref{AH}), this quantity can also be viewed as the ratio of the expected total number of events we observe by $\tau$ and the expected total observation time by $\tau$ when there is no censoring until $\tau.$
In the Epidemiology textbook, the person-time incidence rate is defined by the ratio of the total number of observed events and the total person-time of exposure \citep{rothman2008modern}. When a Poisson model is true (that is, the event time follows an exponential distribution), this is the maximum likelihood estimator for the rate parameter of the Poisson model. However, if the Poisson model is not correct, it will not converge to a population quantity solely related to the event time distribution of $T,$ but one related to both the event time and the censoring time distributions \citep{Uno2023-sm}.  
While the average hazard with survival weight defined by (\ref{AH}) is not a random quantity but a population parameter, it can be interpreted as the average person-time incidence rate of $T$ on $t \in [0,\tau]$ when all $T$ before $\tau$ would have been observed without being censored by the study-specific censoring time. 
From this point forward, the average hazard with survival weight is simply referred to as the average hazard (AH).
Numerical studies conducted by \cite{Uno2023-sm} showed that AH-based tests can be more powerful than the log-rank test and the standard RMST-based tests in detecting delayed treatment effects. 

%------------------------------------------------------ 
% 5. Aims of this paper and structure of this paper 
%------------------------------------------------------ 

In this paper, we seek to gain further power improvement in detecting a delayed treatment effect by combining these two directions of work. 
In Section 2, we propose a long-term average hazard (LT-AH) focusing on the intensity of event occurrence in a later study time window where the survival benefit of immunotherapy appears.
We show nonparametric inference procedures for the ratio of LT-AH and the difference in LT-AH.  
In Section 3, we conduct numerical studies to assess the performance of the proposed approach in finite sample size situations, compared to other methods.
In Section 4, we apply the proposed method to data from the aforementioned randomized study report that compared nivolumab plus ipilimumab with sunitinib in patients with advanced renal cell carcinoma \citep{Motzer2018-ie}. 
Remarks and conclusions are given in Section 5.

%======================
\section{Method}
%======================
\subsection{Long-Term Average Hazard}
Let $T_k$ be a continuous non-negative random variable to denote the event time for group $k \ (k=0,1)$.
Let $C_k$ denote the censoring time for group $k.$ 
Assume that $T_k$ is independent of $C_k.$ 
Let $\left\{ (T_{ki},C_{ki}); \ i=1,\ldots,n_k \right\}$ denote independent copies from $(T_k, C_k).$ 
Let $X_{ki} = \min(T_{ki}, C_{ki})$ and $\Delta_{ki}=I(T_{ki} \le C_{ki}).$
The observable data from group $k$ are then denoted by 
$\left\{ (X_{ki},\Delta_{ki}); \ i=1,\ldots, n_k  \right\}.$ 
We assume $p_k = \lim_{n \rightarrow \infty} n_{k}/n > 0 $ for $k=0,1,$ where $n=n_1+n_0.$ 

Let $h_k(\cdot)$ and $w_k(\cdot)$ be the hazard function for $T_k$ and a continuous nonnegative function, respectively.
A general form of the weighted average of the hazard function over a given time range $[\tau_{1}, \tau_{2}]$ $(0 \le \tau_{1} < \tau_{2})$ is denoted by 
\begin{equation*} 
 \eta_k(\tau_{1}, \tau_{2}) = 
 \frac{ \int_{\tau_{1}}^{\tau_{2}} h_k(u)w_k(u) du}{\int_{\tau_{1}}^{\tau_{2}} w_k(u) du}.
\label{eqn:ltah1}
\end{equation*}

Here, we use the survival function $S_k(t)$ for the weight function $w_k(t)$ as proposed by \cite{Uno2023-sm}.
We assume that $F_k(\tau_{2})> F_k(\tau_{1}) \ge 0$ and $R_k(\tau_{2})-R_k(\tau_{1})>0,$ for $k=0,1,$
where $F_k(\tau) = 1-S_k(\tau)$  is the cumulative incidence probability at $\tau$ and $R_k(\tau) = \int_{0}^{\tau}S_k(u) du$ is the RMST with the truncation time $\tau.$
The average hazard with survival weight over the time range $[\tau_{1}, \tau_{2}]$ is then denoted by 
\begin{equation} 
%\begin{split}
\eta_k(\tau_{1}, \tau_{2}) 
%  = \frac{ \int_{\tau_{1}}^{\tau_{2}} h_k(u)S_k(u) du} 
%   {\int_{\tau_{1}}^{\tau_{2}} S_k(u) du} 
 = \frac{ F_k(\tau_{2}) - F_k(\tau_{1}) } {R_k(\tau_{2}) - R_k(\tau_{1})}   \\
 = \frac{ E\{ I(\tau_1 < T \le \tau_2) \} }
       { E (T \wedge \tau_2) - E (T \wedge \tau_1)}.
\label{eqn:ltah2}
%\end{split}
\end{equation}
The numerator of the right hand side of the equation (\ref{eqn:ltah2})
denotes the probability of having an event between $(\tau_1, \tau_2)$, and the denominator is the expected time of being alive between $(\tau_1, \tau_2).$
Thus, $\eta_k(\tau_{1}, \tau_{2})$ can be interpreted as an average intensity of having an event over the time window $(\tau_1, \tau_2).$ In this paper, we call this quantity the long-term average hazard (LT-AH) hereafter to distinguish it from the AH.

  %------------ 
  % Inference
  %------------ 

A natural non-parametric estimator for the LT-AH is 
\begin{equation} 
 \hat{\eta}_k(\tau_{1}, \tau_{2}) =  \frac{ \hat{F}_k(\tau_{2}) - \hat{F}_k(\tau_{1}) } {\hat{R}_k(\tau_{2}) - \hat{R}_k(\tau_{1})},
\label{eqn:LTAH2}
\end{equation}
where $\hat{F}_k(\tau) = 1- \hat{S}_k(\tau),$ $\hat{R}_k(\tau) = \int_0^{\tau} \hat{S}_k(u) du$, and $\hat{S}_k(\cdot)$ is the Kaplan-Meier estimator for $S_k(\cdot).$
Using the uniform consistency of the Kaplan-Meier estimator \citep{Gill1983-vq}, it can be shown that $\hat{\eta}_k(\tau_{1}, \tau_{2})$ converges in probability to ${\eta}_k(\tau_{1}, \tau_{2})$ as $n_k$ goes to $\infty.$
We consider 
\begin{equation} 
\begin{split}
Q_k &= n_k^{1/2} \left\{\log \hat{\eta}_{k}(\tau_{1}, \tau_{2}) - {\log \eta}_{k}(\tau_{1}, \tau_{2}) \right\} \\ 
    &= n_{k}^{1/2}\left(\log\left\{ \hat{F}_{k}(\tau_{2})-\hat{F}_{k}(\tau_{1})\right\} -\log\left\{ F_{k}(\tau_{2})-F_{k}(\tau_{1})\right\} \right. \\
    & \left. -\left[\log\left\{ \hat{R}_{k}(\tau_{2})-\hat{R}_{k}(\tau_{1})\right\} -\log\left\{ R_{k}(\tau_{2})-R_{k}(\tau_{1})\right\} \right]\right).
\label{eqn:Qk1}
\end{split}
\end{equation}
In Appendix A, we show that $Q_k$ converges weakly to a normal distribution with mean 0 and variance
\begin{equation*}
%\begin{split}
V(Q_k) 
  =\int_0^{\tau_{2}}
\left\{ \frac {{S}_k(\tau_{2})-I(u\le\tau_{1}) {S}_k(\tau_{1}) } {{F}_k(\tau_{2})-{F}_k(\tau_{1})}
  + \frac {\int_u^{\tau_{2}} {S}_k(t)dt - I(u\le\tau_{1}) \int_u^{\tau_{1}} {S}_k(t)dt} {{R}_k(\tau_{2})-{R}_k(\tau_{1})} 
\right\}^{2}\frac{dH_{k}(u)}{G_{k}(u)},
\label{eqn:Var_Qk}
%\end{split}
\end{equation*} 
where $H_k(t)$ is the cumulative hazard function of $T_{k}$ 
and $G_k(t) = \Pr(T_k \wedge C_k \ge t).$ 
We can get an estimate for $V(Q_k)$ by replacing the unknown quantities with their empirical counterparts, as shown below.
\begin{equation*}
%\begin{split}
\hat{V}(Q_{k})  =\int_{0}^{\tau_{2}}\left\{ \frac{\hat{S}_{k}(\tau_{2})-I(u\le\tau_{1})\hat{S}_{k}(\tau_{1})}{\hat{F}_{k}(\tau_{2})-\hat{F}_{k}(\tau_{1})}
 + \frac{\int_{u}^{\tau_{2}}\hat{S}_{k}(t)dt-I(u\le\tau_{1})\int_{u}^{\tau_{1}}\hat{S}_{k}(t)dt}{\hat{R}_{k}(\tau_{2})-\hat{R}_{k}(\tau_{1})}\right\} ^{2}\frac{d\hat{H}_{k}(u)}{\hat{G}_{k}(u)},
\label{eqn:Var_Uk2}
%\end{split}
\end{equation*} 
where $\hat{G}_k(t) =  n^{-1}_k \sum_{i=1}^{n_k} I(X_{ki}\ge t),$ 
and $\hat{H}_k(\cdot)$ is the Nelson-Aalen estimator for $H_k(\cdot)$ for group $k.$
Using these results, an $(1-\alpha)$ asymptotic confidence interval (CI) for the LT-AH in group $k$ will be derived by 
$$\exp \left\{ \log \hat{\eta}_k(\tau_{1}, \tau_{2}) \pm z_{1-\alpha/2} \sqrt{\hat{V}(Q_k)/n_k} \right\},$$  where $z_{1-\alpha/2}$ is the $(1-\alpha/2)\times 100 $-percentile of the standard normal distribution.

%--------------------------------------------------------------------------------------------------------------------
\subsection{Between-group contrast measures derived from LT-AH}
Using the group-specific LT-AH's from two groups, we can summarize the magnitude of the treatment effect in both absolute difference and relative terms.
In this section, we describe the results regarding the inference of the ratio of LT-AH and the difference in LT-AH. 
%--------------------------------------------------------------------------------------------------------------------
\subsubsection{Ratio of LT-AH}
We estimate the ratio of LT-AH 
\begin{equation*}
\theta(\tau_{1}, \tau_{2})  = \frac{ \eta_1(\tau_{1}, \tau_{2}) } {\eta_0(\tau_{1}, \tau_{2})} = 
\frac{ F_1(\tau_{2})-F_1(\tau_{1}) } {F_0(\tau_{2})-F_0(\tau_{1}) } \frac{ R_0(\tau_{2})-R_0(\tau_{1}) } {R_1(\tau_{2})-R_1(\tau_{1}) }
\label{eqn:ratio_ltah}
\end{equation*}
by $$\hat{\theta}(\tau_{1}, \tau_{2})  =  \frac{\hat{\eta}_1(\tau_{1}, \tau_{2})}{\hat{\eta}_0(\tau_{1}, \tau_{2})}.$$
For the inference of $\theta(\tau_{1}, \tau_{2})$, we consider the asymptotic distribution of 
\begin{equation*}
\begin{split}
& n^{1/2} \left\{ \log \hat{\theta}(\tau_{1}, \tau_{2}) - \log {\theta}(\tau_{1}, \tau_{2}) \right\} \\
 & = n^{1/2} \left\{ \log\hat{\eta}_{1}(\tau_{1}, \tau_{2}) - \log{\eta}_{1}(\tau_{1}, \tau_{2}) \right\} 
  - n^{1/2} \left\{ \log\hat{\eta}_{0}(\tau_{1}, \tau_{2}) - \log{\eta}_{0}(\tau_{1}, \tau_{2}) \right\}.
\end{split}
\end{equation*}
From the results regarding $Q_k$ described in Section 2.1, 
it is shown that \\ $n^{1/2} \left\{ \log \hat{\theta}(\tau_{1}, \tau_{2}) - \log {\theta}(\tau_{1}, \tau_{2}) \right\} 
$ converges weakly to a normal distribution with mean zero and variance  
$ p^{-1}_1{V}(Q_1) + p^{-1}_0{V}(Q_0).$
The variance can be estimated by 
$\hat{p}_1^{-1}\hat{V}(Q_1) + \hat{p}_0^{-1}\hat{V}(Q_0),$
where $\hat{p}_k = n_k/n,$ for $k=0,1.$ 
Therefore, an $(1-\alpha)$ asymptotic CI for $\theta(\tau)$ is 
\begin{equation}
\exp \left\{ \log \hat{\theta}(\tau_{1}, \tau_{2}) \pm z_{1-\alpha/2} 
\sqrt{n_1^{-1}\hat{V}(Q_1) + n_0^{-1}\hat{V}(Q_0)}
\right\}.
\label{eqn:CI}
\end{equation}
For testing the null hypothesis, $\theta(\tau_{1}, \tau_{2}) = 1,$
\begin{equation}
\log\hat{\theta}(\tau_{1}, \tau_{2})/ 
\sqrt{n_1^{-1}\hat{V}(Q_1) + n_0^{-1}\hat{V}(Q_0)}
\label{eqn:test}
\end{equation}
is used as the test statistic, which asymptotically follows the standard normal distribution under the null hypothesis.

%--------------------------------------------------------------------------------------------------------------------
\subsubsection{Difference in LT-AH}
We estimate the difference in LT-AH 
\begin{equation*}
\xi(\tau_{1}, \tau_{2})  = \eta_1(\tau_{1}, \tau_{2}) - \eta_0(\tau_{1}, \tau_{2})  = \frac{F_1(\tau_{2})-F_1(\tau_{1})} {R_1(\tau_{2})-R_1(\tau_{1})}-\frac{F_0(\tau_{2})-F_0(\tau_{1})} {R_0(\tau_{2})-R_0(\tau_{1})}
\label{eqn:diff_ltah}
\end{equation*}
by $$ \hat{\xi}(\tau_{1}, \tau_{2})=\hat{\eta}_1(\tau_{1}, \tau_{2}) - \hat{\eta}_0(\tau_{1}, \tau_{2}).$$
For hypothesis testing and interval estimation, we consider the following asymptotic distribution
\begin{equation}
\begin{split}
& n^{1/2} \left\{ \hat{\xi}(\tau_{1}, \tau_{2}) - {\xi}(\tau_{1}, \tau_{2}) \right\} \\
& =   n^{1/2} \left\{ \hat{\eta}_{1}(\tau_{1}, \tau_{2}) - {\eta}_{1}(\tau_{1}, \tau_{2}) \right\}  - n^{1/2} \left\{ \hat{\eta}_{0}(\tau_{1}, \tau_{2}) - {\eta}_{0}(\tau_{1}, \tau_{2}) \right\}.
\label{dist_xi_12}
\end{split}
\end{equation}
In Appendix B, it is shown that 
\begin{equation} 
%\begin{split}
U_k = n_k^{1/2} \left\{\hat{\eta}_{k}(\tau_{1}, \tau_{2}) - {\eta}_{k}(\tau_{1}, \tau_{2}) \right\}
 = n_k^{1/2} \left\{\frac{ \hat{F}_k(\tau_{2}) - \hat{F}_k(\tau_{1}) } { \hat{R}_k(\tau_{2}) - \hat{R}_k(\tau_{1})} - \frac{ F_k(\tau_{2}) - F_k(\tau_{1}) } {R_k(\tau_{2}) - R_k(\tau_{1})}\right\}
\label{eqn:Uk1}
%\end{split}
\end{equation}
converges weakly to a normal distribution with mean 0 and variance
\begin{equation*}
\begin{split}
& 
V(U_k) = 
\int_0^{\tau_{2}} 
\left[ 
\frac 
{{S}_k(\tau_{2})-I(u\le \tau_{1}){S}_k(\tau_{1})} {{R}_k(\tau_{2}) - {R}_k(\tau_{1})} \right. \\
& + 
\left. \frac
{ {F}_k(\tau_{2})- {F}_k(\tau_{1}) }
{ \left\{ {R}_k(\tau_{2})- {R}_k(\tau_{1}) \right\}^2}
\left\{ \int_u^{\tau_{2}} {S}_k(t)dt - I(u \le \tau_{1}) \int_u^{\tau_{1}} {S}_k(t)dt \right\} 
\right]^2
\frac{dH_{k}(u)}{G_{k}(u)}.
\end{split}
\end{equation*} 
Applying this result to the equation (\ref{dist_xi_12}), it is shown that $n^{1/2} \left\{ \hat{\xi}(\tau_1,\tau_2) - {\xi}(\tau_1,\tau_2) \right\}$ 
converges weakly to a normal distribution with mean 0 and variance
$ p^{-1}_1{V}(U_1) + p^{-1}_0{V}(U_0)$. 
This can be estimated by replacing the unknown quantities with their empirical counterparts, $ \hat{p}^{-1}_1\hat{V}(U_1) + \hat{p}^{-1}_0\hat{V}(U_0)$. Therefore, an $(1-\alpha)$ asymptotic CI for $\xi(\tau_1,\tau_2)$ is given by
\begin{equation} 
\hat{\xi}(\tau_1,\tau_2) \pm z_{1-\alpha/2} \sqrt{n^{-1}_1\hat{V}(U_1) + n^{-1}_0\hat{V}(U_0)}.
\label{eqn:CI_DAH}
\end{equation} 
For testing the null hypothesis that there is no difference in LT-AH with time window ($\tau_{1}$,$\tau_{2}$) between two groups (i.e., $\xi(\tau_1,\tau_2)=0$), we will use
\begin{equation}
\hat{\xi}(\tau_1,\tau_2) / \sqrt{n^{-1}_1\hat{V}(U_1) + n^{-1}_0\hat{V}(U_0)} 
\label{eqn:test_DAH}
\end{equation} 
as the test statistic, which asymptotically follows the standard normal distribution under the null.

%==========================
\section{Numerical Studies}
%==========================
We performed extensive numerical studies and evaluated finite sample properties of the proposed asymptotic 0.95 CIs for difference in LT-AH and ratio of LT-AH and asymptotic tests for no treatment effect 
(i.e., $\xi(\tau_1,\tau_2)=0$ and $\theta(\tau_1,\tau_2)=1,$ respectively).

\subsection{Configurations}
%---
Figure 2 shows five different patterns for the event time distribution between two groups considered in our numerical studies. 
Pattern 2(A) denotes a no difference scenario. 
This was used to confirm that the sizes of the tests we examined in the numerical studies were at the conventional nominal level (i.e., 5.0\%). 
Pattern 2(B) denotes a proportional hazards difference. 
For delayed difference scenarios, we considered three patterns: 2(C) to 2(E). 
Specifically, two survival curves were identical up to time 2 and then (I) diverging in Pattern 2(C), (II) almost parallel in Pattern 2(D), and (III) converging in Pattern 2(E) as heading to the end of the follow-up. 
For all patterns, the event time in group 0, $T_0,$ was generated from the Weibull distribution with parameters (shape, scale)=(1,10). 
For Pattern 2(B), the event time in group 1, $T_1,$  was generated from the Weibull distribution with parameters (shape, scale)=(1, 12.5). 
For Patterns 2(C) to 2(E), we first generated a random number, $U$, from a standard uniform distribution. We then transformed it to the event time in group 1 by 
$T_{1} = S_{1}^{-1}(U),$ where $S_{1}(t)$ is the survival function in group 1 as presented with the solid line in Figures 2(C) to 2(E), respectively.

%%%%%%%%%%%%%%%%%%%%%%%%%%%%%%%%%%%%
% Figure 2: Survival time distribution used in the numerical study
%%%%%%%%%%%%%%%%%%%%%%%%%%%%%%%%%%%%
%\clearpage
%\newpage
\begin{figure}
\label{figure2}
\centering
\includegraphics[scale=0.7 ]{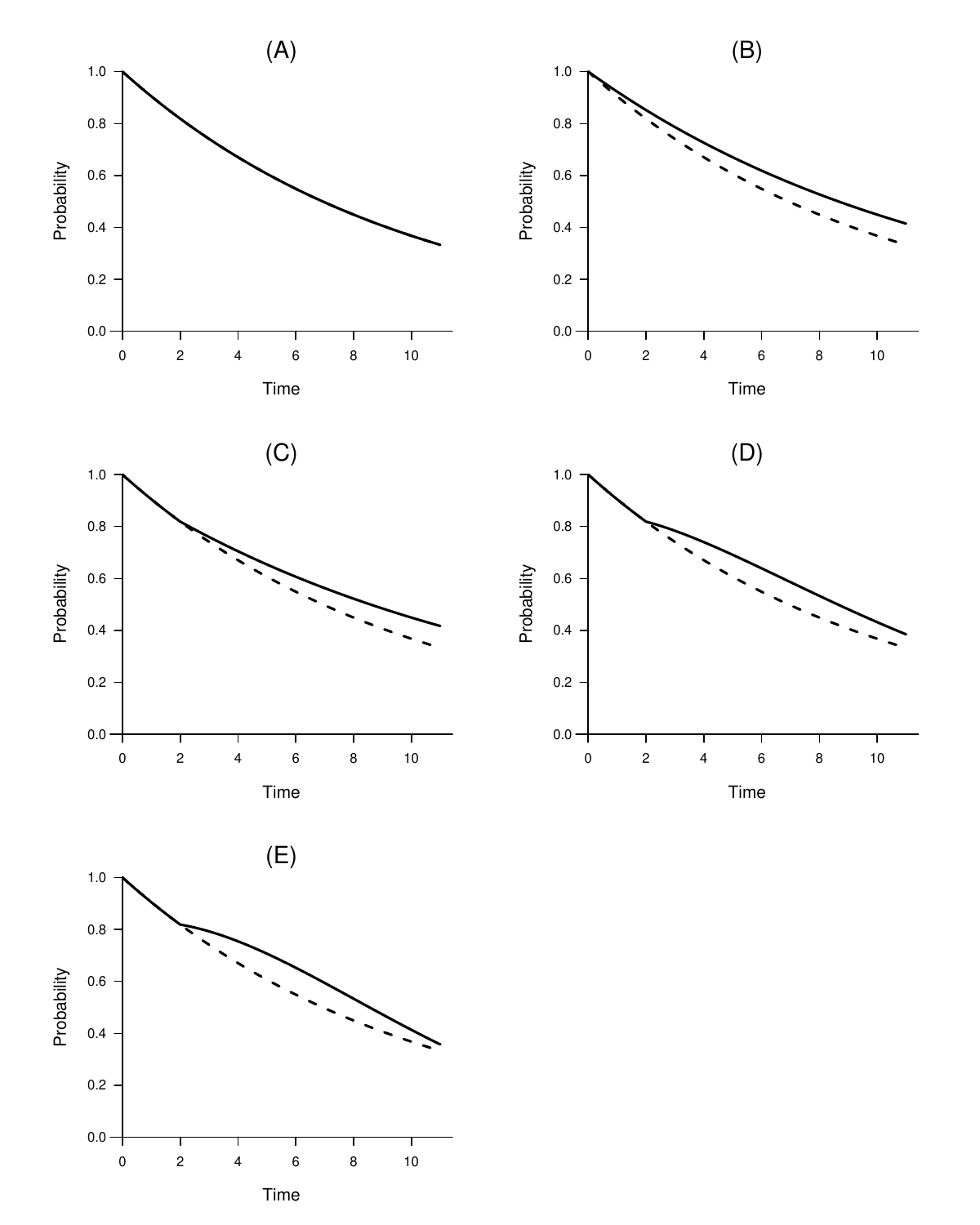}
\caption{Survival functions of event time distributions of the treatment group (solid line) and control group (dashed line) used in the numerical studies. (A), no difference; (B), proportional hazards difference; (C), delayed difference I; (D), delayed difference II; (E), delayed difference III.}
\end{figure}

Figure 3 shows three patterns of censoring time distributions considered in our numerical studies: no censoring pattern (i), light censoring pattern (ii), and moderate censoring pattern (iii). We used the Weibull distributions with parameters (shape, scale)=(3.871, 14.189) for Pattern (ii), and (shape, scale)=(2.818, 10.233) for Pattern (iii). 
The administrative censoring at time 10 was applied to all three censoring time distribution patterns. 
The censoring time distribution was not group-specific but study-specific in our numerical studies. As such, a total of 15 combinations of difference in the event time distribution and censoring time distribution were investigated. 

%%%%%%%%%%%%%%%%%%%%%%%%%%%%%%%%%%%%
% Figure 3: Censoring time distribution used in the numerical study
%%%%%%%%%%%%%%%%%%%%%%%%%%%%%%%%%%%%
%\clearpage
%\newpage
\begin{figure}
\label{figure3}
\centering
\includegraphics[scale=0.8 ]{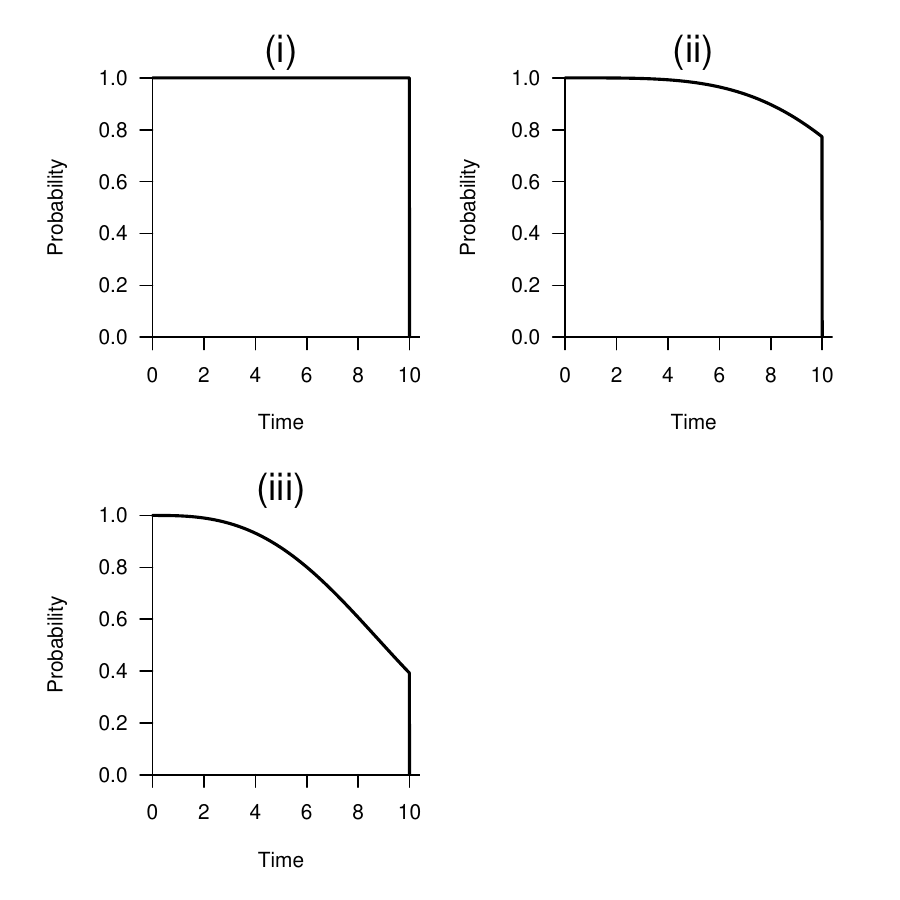}
\caption{Survival functions of censoring time distributions used in the numerical studies. (i), no censoring; (ii), light censoring; (iii), moderate censoring.}
\end{figure}

%----
From each of the 15 combinations, we generated $n_k$ pairs of data points for each group $\{(T_{ki},C_{ki}); \ k=0, 1,\ i=1,\ldots,n_{k}\}$ independently, where the censoring time was independent of the event time.  We then derived the observable data $\left\{ (X_{ki},\Delta_{ki});\ k=0,1,\ i=1,\ldots,n_{k}\right\}$, where $X_{ki}=T_{ki} \wedge C_{ki}$ and $\Delta_{ki}=I(T_{ki}\le C_{ki}).$
We then calculated the difference in LT-AH and the ratio of LT-AH, and the corresponding 0.95 CIs, using (\ref{eqn:CI}) and (\ref{eqn:CI_DAH}), respectively.
The asymptotic tests based on the difference in LT-AH and the ratio of LT-AH were also performed, using the test statistics  (\ref{eqn:test}) and (\ref{eqn:test_DAH}), respectively, at two-sided 0.05 $\alpha$ level. 

Repeating this process 5,000 times, we assessed the empirical bias of the proposed point estimators for difference in LT-AH and ratio of LT-AH, the empirical coverage probability of the 0.95 CIs, and the average length of 0.95 CIs. 
We also evaluated the empirical size and power of the asymptotic tests. As comparators, we included the score test based on Cox's method, and tests based on the standard RMST, LT-RMST, and AH. 
For the standard RMST and AH, the time window [0, 10] was considered. The time range [2, 10] was used for the analysis based on LT-AH and LT-RMST. 
We considered three kinds of sample size for each scenario: $n_{k}=100, 200,$ and $300$ $(k=0,1).$

\subsection{Results}
%Coverage
Table 1 summarizes the performance of the difference in LT-AH and ratio of LT-AH with a sample size of 100 per arm. 
Regarding the difference in LT-AH, the absolute value of the empirical bias of the proposed estimator was less than 0.0005 in all scenarios. The empirical coverage probabilities were close to the nominal level of 0.95 (ranging from 0.945 to 0.953). 
The average length of the CI increased slightly as the censoring increased, as expected. 
The results of bias and coverage with sample sizes of 200 and 300 per arm were almost identical to those of Table 1 (data not shown). 

Regarding the ratio of LT-AH, the empirical bias ranged from 0.015 to 0.034 with sample size 100 per arm (Table 1). The bias with moderate censoring was relatively greater than that with no or light censoring. 
The empirical coverage probabilities were close to the nominal level (ranging from 0.947 to 0.952). 
As the sample size increased, the empirical bias decreased. 
Specifically, the bias ranged from 0.006 to 0.017 with sample size 200 per arm and from 0.004 to 0.012 with sample size 300 per arm. Coverage probabilities with sample sizes 200 and 300 per arm were almost identical to those with sample size 100 per arm (data not shown).

%%%%%%%%%%%%%%%%%%%%%%%%%%%%%%%%%%%%%%%%%%%%%%%%
%% Table 1:  Coverage (n=100 per arm)%%
%%%%%%%%%%%%%%%%%%%%%%%%%%%%%%%%%%%%%%%%%%%%%%%%
%\clearpage
%\newpage
\begin{table}
\label{table1}
\caption{Performance of the difference and ratio of long-term average hazard with survival weight with sample size 100 per arm.}
\centering

%---param_num=10.1, 10.2, 10.3
\begin{tabular}{lcccccccc}
\hline
&\multicolumn{4}{c}{\textbf{Difference in LT-AH}}&\multicolumn{4}{c}{\textbf{Ratio of LT-AH}}\tabularnewline
\hline
 & \textbf{True} & \textbf{Bias}  & \textbf{CP}  & \textbf{AL} & \textbf{True} & \textbf{Bias}  & \textbf{CP}  & \textbf{AL} \tabularnewline
\hline
\textbf{Censoring} & &  & & & & & &  \tabularnewline
\hline
\multicolumn{9}{c}{Event time distribution pattern: No difference}\tabularnewline
\hline
  None        & 0.000 & -0.000 & 0.953 & 0.082 & 1.000 & 0.021 & 0.951 & 0.865 \\ 
  Light         & 0.000 & 0.000 & 0.953 & 0.085 & 1.000 & 0.027 & 0.951 & 0.894 \\ 
  Moderate & 0.000 & 0.000 & 0.948 & 0.094 & 1.000 & 0.034 & 0.948 & 1.015 \\ 
\hline

%---param_num=20.1, 20.2, 20.3
\multicolumn{9}{c}{Event time distribution pattern: PH difference}\tabularnewline
\hline 
  None       & -0.020 & -0.000 & 0.953 & 0.076 & 0.800 & 0.016 & 0.950 & 0.712 \\ 
  Light        & -0.020 & 0.000 & 0.949 & 0.078 & 0.800 & 0.021 & 0.947 & 0.737 \\ 
  Moderate & -0.020 & 0.000 & 0.946 & 0.088 & 0.800 & 0.027 & 0.948 & 0.840 \\ 
\hline

%---param_num=31.1, 31.2, 31.3
\multicolumn{9}{c}{Event time distribution pattern: Delayed difference I}\tabularnewline
\hline 
  None        & -0.025 & -0.000 & 0.952 & 0.076 & 0.750 & 0.015 & 0.952 & 0.684 \\  
  Light         & -0.025 & 0.000 & 0.952 & 0.078 & 0.750 & 0.019 & 0.949 & 0.709 \\ 
  Moderate & -0.025 & 0.000 & 0.947 & 0.087 & 0.750 & 0.026 & 0.949 & 0.808 \\ 
\hline

%---param_num=32.1, 32.2, 32.3
\multicolumn{9}{c}{Event time distribution pattern: Delayed difference II}\tabularnewline
\hline
  None        & -0.024 & -0.000 & 0.951 & 0.074 & 0.762 & 0.015 & 0.951 & 0.666 \\  
  Light        & -0.024 & -0.000 & 0.949 & 0.076 & 0.762 & 0.019 & 0.948 & 0.693 \\ 
  Moderate & -0.024 & 0.000 & 0.947 & 0.086 & 0.762 & 0.025 & 0.946 & 0.800 \\ 
\hline

%---param_num=33.1, 33.2, 33.3
\multicolumn{9}{c}{Event time distribution pattern: Delayed difference III}\tabularnewline
\hline
  None       & -0.021 & -0.000 & 0.949 & 0.073 & 0.791 & 0.015 & 0.951 & 0.673 \\ 
  Light        & -0.021 & -0.000 & 0.949 & 0.076 & 0.791 & 0.019 & 0.950 & 0.701 \\ 
  Moderate & -0.021 & 0.000 & 0.945 & 0.086 & 0.791 & 0.025 & 0.947 & 0.814 \\ 
\hline
\end{tabular}

\centering

\bigskip{}
\bigskip{}
\begin{minipage}{390pt}
Event time distribution pattern: 2(A), no difference; 2(B), PH difference; 2(C), delayed difference I; 2(D), delayed difference II; 2E, delayed difference III (see Figure 2). 
Censoring time distribution pattern: 3(i), no censoring; 3(ii), light censoring; 3(iii), moderate censoring (see Figure 3).
\newline
Abbreviations: 
LT-AH, long-term average hazard;
PH, proportional hazards; 
Censoring, censoring time distribution pattern; 
True, the true value;
Bias, the empirical bias (estimate minus true value); 
CP, the empirical coverage probability of the 0.95 confidence interval; 
AL, the average length of the 0.95 confidence interval.

\end{minipage}
\end{table}

%--Type 1 errors
The size and power of the asymptotic tests with a sample size of 200 per arm were summarized in Table 2. Since the number of iterations was 5000 and the true type I error rate is 5.0\%, the empirical type I error rate should be within 4.4\% to 5.6\% with 95\% chance. According to the pattern of no difference (Figure 2(A)), the type I empirical error rates of all tests were within this range for all censoring patterns.

%--Power
%--PH pattern
The power of each test was also summarized in Table 2. Under the proportional hazards difference pattern (Figure 2(B)), Cox's hazard ratio test showed the highest power as indicated by the theories. 
The test based on AH difference showed power comparable to Cox's hazard ratio, followed by the RMST-based and LT-RMST-based tests. The LT-RMST-based test showed higher power than the RMST-based test. As expected given its focus on a subset of events, the LT-AH-based test showed lower power than the AH-based test under the proportional hazards pattern. 
The same trends were observed in the three censoring patterns. The power decreased with increased censoring for each test.

%--Delayed differences I to III
For the delayed difference scenarios (Figures 2(C) to 2(E)), the long-term versions of the AH-based and RMST-based tests showed higher power than their respective counterparts in all three delayed difference patterns from I to III, which supported using the long-term version when a delayed difference is expected.
However, the choice between the LT-AH-based test and the LT-RMST-based test may depend on the delayed difference pattern.
Specifically, in the delayed difference pattern I (i.e., a proportional hazards difference after separation; see Figure 2(C)), the LT-AH-based test was more powerful than the LT-RMST-based test. However, the results were the opposite in the delayed difference pattern III (i.e., two survival curves do not diverge but converge after separation; see Figure 2(E)).
In the delayed difference pattern II (i.e., separation of two survival curves after the separation point looks parallel; see Figure 2(D)), the LT-AH-based and LT-RMST-based tests showed similar power, but the superiority of these two tests also depended on the censoring pattern.

These studies also showed that the LT-AH-based test is superior to Cox's hazard ratio test in delayed difference patterns when censoring is none or light. For example, under light censoring, the power of Cox's hazard ratio test versus the LT-AH-based test was (0.316 vs. 0.429), 
(0.332 vs. 0.407), and (0.288 vs. 0.330) for the delayed difference pattern I to III, respectively. 
On the other hand, the superiority of the LT-RMST-based test over Cox's hazard ratio test depended on the pattern of delayed difference. Specifically, Cox's hazard ratio test was more powerful than the LT-RMST-based test for the delayed difference pattern I (0.333 vs. 0.247; no censoring), but the results were the opposite in the delayed difference patterns II (0.320 vs. 0.425; no censoring) and III (0.262 vs. 0.485; no censoring). 

%----
We carried out power comparisons using ratios of AH, LT-AH, RMST and LT-RMST in parallel to the differences in these metrics, but we did not include the results here because they were almost identical.
Also, we had the same findings from the results with sample sizes 100 and 300 per arm (data not shown).

In summary, when delayed onset of treatment benefit is expected, LT-AH or LT-RMST is recommended. They will offer higher power than the standard AH-based test and RMST-based test, respectively. 
Especially when censoring is light, the LT-AH-based test will have a higher power than Cox's hazard ratio test under many delayed difference patterns. On the other hand, LT-RMST-based test can be more powerful or less powerful than Cox's hazard ratio test, depending on the delayed difference pattern. 
LT-AH would be preferable to LT-RMST when two survival curves keep diverging after the separation time point (Pattern I; Figure 2(C)). Conversely, LT-RMST would be preferable to LT-AH if two survival curves diverge after the separation time point and then converged as they head to the end of the follow-up (Pattern III; Figure 2(E)).

%%%%%%%%%%%%%%%%%%%%%%%%%%%%%%%%%%%%%%%%%%%%%
%% Table 2:  Power (n=200 per arm)%% Difference
%%%%%%%%%%%%%%%%%%%%%%%%%%%%%%%%%%%%%%%%%%%%%
%\clearpage
%\newpage

\begin{table}[h]
\label{table2}
\caption{Size and power of tests based on Cox's hazard ratio, difference in average hazard, difference in long-term average hazard, difference in restricted mean survival time, and difference in long-term restricted mean survival time with sample size 200 per arm.}
\centering
%\small

\begin{tabular}{lcccccc}

\hline
 &  & \multicolumn{4}{c}{\textbf{Difference in}} \\
\cline{3-6}

 & \textbf{Cox's HR} & \textbf{AH} & \textbf{LT-AH}  & \textbf{RMST}  & \textbf{LT-RMST} \tabularnewline
\hline
\\
\textbf{Censoring} & \multicolumn{5}{c}{\textbf{Size of tests}}\tabularnewline
\hline
& \multicolumn{5}{c}{Event time distribution pattern: No difference}\tabularnewline
\hline
%---param_num=10.1, 10.2, 10.3
  None       & 0.047 & 0.047 & 0.044 & 0.048 & 0.048 \\ 
  Light        & 0.052 & 0.052 & 0.048 & 0.052 & 0.055 \\ 
  Moderate & 0.049 & 0.047 & 0.049 & 0.051 & 0.052 \\ 
  \hline
\\
\textbf{Censoring} & \multicolumn{5}{c}{\textbf{Power of tests}}\tabularnewline
\hline
& \multicolumn{5}{c}{Event time distribution pattern: PH difference}\tabularnewline
\hline 
%---param_num=20.1, 20.2, 20.3
  None       & 0.403 & 0.402 & 0.308 & 0.355 & 0.370 \\ 
  Light        & 0.386 & 0.383 & 0.291 & 0.352 & 0.362 \\ 
  Moderate & 0.350 & 0.336 & 0.236 & 0.339 & 0.348 \\ 
\hline
& \multicolumn{5}{c}{Event time distribution pattern: Delayed difference I}\tabularnewline
\hline 
%---param_num=31.1, 31.2, 31.3
  None       & 0.333 & 0.337 & 0.453 & 0.217 & 0.247 \\ 
  Light        & 0.316 & 0.325 & 0.429 & 0.215 & 0.246 \\ 
  Moderate & 0.254 & 0.281 & 0.347 & 0.202 & 0.235 \\ 
\hline
& \multicolumn{5}{c}{Event time distribution pattern: Delayed difference II}\tabularnewline
\hline 
%---param_num=32.1, 32.2, 32.3
  None       & 0.320 & 0.322 & 0.434 & 0.362 & 0.425 \\ 
  Light        & 0.332 & 0.313 & 0.407 & 0.360 & 0.417 \\ 
  Moderate & 0.339 & 0.265 & 0.323 & 0.347 & 0.398 \\ 
\hline
& \multicolumn{5}{c}{Event time distribution pattern: Delayed difference III}\tabularnewline
\hline 
%---param_num=33.1, 33.2, 33.3
  None       & 0.262 & 0.258 & 0.352 & 0.417 & 0.485 \\ 
  Light        & 0.288 & 0.250 & 0.330 & 0.409 & 0.472 \\ 
  Moderate & 0.335 & 0.214 & 0.253 & 0.395 & 0.454 \\ 
\hline
\end{tabular}

%\bigskip{}
%\bigskip{}
\vspace{0.2cm}
\begin{minipage}{290pt}
Event time distribution pattern: 2(A), no difference; 2(B), PH difference; 2(C), delayed difference I; 2(D), delayed difference II; 2(E), delayed difference III (see Figure 2). 
Censoring time distribution pattern: 3(i), no censoring; 3(ii), light censoring; 3(iii), moderate censoring (see Figure 3).
\newline
Abbreviations: 
HR, hazard ratio;
%Diff, difference;
AH, average hazard;
LT-AH, long-term average hazard;
RMST, restricted mean survival time;
LT-RMST, long-term restricted mean survival time;
PH, proportional hazards; 
Censoring, censoring time distribution pattern.
\end{minipage}
\end{table}

%======================
\section{Example}
%======================
To illustrate the proposed method, we used data from the CheckMate 214 study, a recently conducted randomized clinical trial for assessing immunotherapy in patients with previously untreated clear-cell advanced renal cell carcinoma. Figure 1(A) shows the Kaplan-Meier curves for PFS comparing nivolumab plus ipilimumab with sunitinib, where we reconstructed patient-level PFS data \citep{Guyot:2012kt} from the study results reported by \cite{Motzer2018-ie}. 
The estimated hazard ratio was 0.82 (0.95 CI: 0.68 to 0.99; p-value=0.037), favoring the nivolumab plus ipilimumab arm. 
The estimated survival curve for the nivolumab plus ipilimumab arm was almost identical to that for the sunitinib up to 7 months (Figure 1(A)), and demonstrated a delayed difference pattern in two survival curves. This suggests that the nivolumab plus ipilimumab arm might benefit PFS relatively long-term. However, the analysis results of the hazard-ratio-based test/estimation approach do not address the long-term treatment effect of the immunotherapy. 
Therefore, we used the proposed LT-AH test/estimation approach to capture the long-term benefit, focusing on a time window [7, 21] (months) as illustrated in the shaded areas in Figure 1(B).

The difference in LT-AH on [7, 21] (months) was -0.023 (with LT-AH values of 0.028 for the nivolumab and ipilimumab group and 0.051 for the sunitinib group) (Table 3). This can be interpreted as, on average, that immunotherapy reduces the event rate by 2.3 events per 100 person-months compared to sunitinib (0.95 CI: -0.037 to -0.008; p=0.002) within the study time window between 7 and 21 months.  The ratio of LT-AH was 0.553 (0.95 CI: 0.387 to 0.791; p=0.001).

As a reference, we also applied the LT-RMST test/estimation approach. The difference in RMST over [7, 21] (months) was 1.2 months (with RMST values of 6.7 for the nivolumab and ipilimumab group and 5.5 for the sunitinib group). That is, PFS probability among future patients receiving the immunotherapy would be a mean of 6.7 months from months 7 to 21, which is 1.2 months longer than that among patients treated with sunitinib (0.95 CI: 0.2 to 2.1; p-value=0.017) (Table 3).
The p-value of the LT-AH test for difference was smaller than that of the LT-RMST (0.002 vs. 0.017). This is because the observed difference pattern was similar to the pattern I (Figure 2(C)) used in our numerical studies (i.e., a delayed difference pattern with two survival curves keep diverging after the separation time point). 

Also, as a reference, we applied the standard AH-based and standard RMST-based methods with the time window [0, 21 months] (Table 3). The p-values of the LT-AH and LT-RMST comparison were lower than those of the standard AH-based and standard RMST-based tests, respectively. 
These show an advantage of using the long-term versions of these metrics in cases where the difference between groups was considered as noise in the early time window. 

%%%%%%%%%%%%%%%%%%%%%%%%%%%%%%%%%%%%
%% Table 3:  Example%%
%%%%%%%%%%%%%%%%%%%%%%%%%%%%%%%%%%%%
%\clearpage
%\newpage
%\setlength{\rotFPtop}{0pt plus 1fil}

%\begin{table}
\begin{sidewaystable}
\label{table3}
\centering
\caption{Estimated restricted mean survival times and average hazards with time rage $[\tau_{1},\tau_{2}]$ for treatment group (nivolumab plus ipilimumab) and control group (sunitinib) with the data reconstructed from the publication of the CheckMate 214 study.}

%\scalebox{0.9}{
\begin{tabular}{lcllll}
  \hline
 & $[\tau_{1},\tau_{2}]$ & & & &  \\ 
 & (month) & Treatment (0.95 CI) & Control (0.95 CI) & Difference$^*$ (0.95 CI; p-value) & Ratio$^{**}$ (0.95 CI; p-value) \\ 

  \hline
   LT-AH   & [7, 21] & 0.028 (0.022 to 0.037) & 0.051 (0.040 to 0.065) & -0.023 (-0.037 to -0.008; 0.002) & 0.553 (0.387 to 0.791; 0.001) \\ 
   AH      & [0, 21] & 0.049 (0.042 to 0.057) & 0.066 (0.057 to 0.076) & -0.017 (-0.029 to -0.005; 0.006) & 0.747 (0.608 to 0.917; 0.005) \\ 
   LT-RMST  & [7, 21] & 6.7 (5.9 to 7.3) & 5.5 (4.8 to 6.1) & 1.2 (0.2 to 2.1; 0.017) & 1.2 (1.0 to 1.4; 0.017) \\ 
   RMST & [0, 21] & 12.2 (11.4 to 13.0) & 11.0 (10.2 to 11.8) & 1.2 (0.0 to 2.4; 0.043) & 1.1 (1.0 to 1.2; 0.041) \\  
   \hline
\end{tabular}
%}
%------------------------
\centering

\vspace{0.2cm}
\begin{minipage}{560pt}
Abbreviations: LT-AH, long-term average hazard; AH, average hazard; LT-RMST, long-term restricted mean survival time; RMST, restricted mean survival time; CI, confidence interval. \\
* Difference: Treatment $-$ Control. A value below 0 is in favor of the treatment group for AH, and that above 0 is in favor of the treatment group for RMST. \\
** Ratio: Treatment/Control. A value below 1 is in favor of the treatment group for AH, and that above 1 is in favor of the treatment group for RMST. \\
\end{minipage}
%\end{table}
\end{sidewaystable}

%======================
\section{Remarks}
%======================
In this paper extending the AH-based approach proposed by \cite{Uno2023-sm}, we proposed a new test/estimation approach using LT-AH to focus on quantifying the long-term treatment benefit of time-to-event outcome. 
The proposed LT-AH-based approach will be particularly useful in some randomized clinical trials examining new therapies with potential delayed treatment effects, such as immunotherapy.
Our numerical studies showed that LT-AH-based tests are more powerful than AH-based tests in detecting a delayed treatment effect. However, the LT-RMST-based tests demonstrated comparable performance. The superiority of these two long-term focused approaches depends on the pattern of difference in the two underlying survival functions after the separation time point. 
Specifically, the strength of LT-AH is to detect a delayed difference by which two survival curves keep diverging after the separation time point. On the other hand, the LT-RMST is more powerful in detecting such a delayed difference pattern that the difference in two survival curves after the separation time point is diminishing as the time elapses. 

In addition to the power advantage in delayed difference scenarios, the proposed LT-AH approach can summarize the magnitude of the treatment effect in both absolute difference and relative terms using ``hazard'' (i.e., difference in LT-AH and ratio of LT-AH), meeting guideline recommendations and practical needs (\citeauthor{Cobos-Carbo2011-wz}, 2011;  \citeauthor{annals}, 2024). Unlike Cox's hazard ratio, these estimates are robust and free from the study-specific censoring time distribution. 
The proposed LT-AH approach can be a useful alternative to the conventional log-rank/hazard-ratio test/estimation approach when the delayed onset of treatment benefit on time-to-event outcomes is expected.

% How to choose tau1 and tau2.
A practical issue with the proposed method is how to specify $[\tau_{1}, \tau_{2}]$. Similar to the LT-RMST-based methods, we recommend that this time window be selected based on clinical considerations \citep{Horiguchi:2018dk}. 
Specifically, if the proposed method is used in confirmatory clinical trials, either a specific time window or a specific rule to determine the time window should be prespecified in the study protocol.
The test results depend on the choice of time window, especially in the presence of a crossing hazard. When the treatment benefit is only delayed, 
$\tau_{1}$ may be selected as the separation time point of two observed survival curves. 
We should ensure that the size of the risk set at $\tau_{2}$ is large enough for the large sample approximation. In practice, sensitivity analysis with several windows would also be encouraged to better understand the treatment effect profile.

% LT-AH and the landmark analysis
Similarly to the piecewise log-rank test \citep{Xu:2017dz} introduced in Section 1, the LT-AH proposed in this paper is equivalent to a version of landmark analysis based on AH (see Appendix C). In an ideal application scenario, survival curves of the two treatment groups are identical up to the landmark time point, and the landmark analysis effectively compares the overall survival profile.  
When two survival curves differ before the landmark time point, 
we emphasize the need to exercise caution with the general limitations of the landmark analysis \citep{Dafni2011-fa}. Especially due to the potential violation of the intention-to-treat principle, a causal interpretation of the analysis results regarding the effect of the treatment on survival time would be almost impossible.   

%Cross-survival case is out of the scope of this paper
In this paper, we restricted delayed difference patterns to the cases where two survival functions were almost identical at the early study time. If two survival functions before the separation time point were not similar, the use of the long-term version of AH or RMST might not be appropriate. For this, we did not include so-called cross-survival cases, in which the Kaplan-Meier curves cross during follow-up, in our considerations or numerical studies. As discussed elsewhere \citep{Horiguchi2023-ie}, for cross-survival scenarios, any single metric would be insufficient to inform which group is superior. Additional assessments using multiple metrics would be required. 

% R package
The software for the implementation of the proposed method (LT-AH) is currently R. 
The latest version of the {\it survAH} R package is available from a repository on the GitHub page of the corresponding author \\ (https://github.com/uno1lab/survAH).

\vspace{1cm}
{\Large{\bf{Appendix}}}
\appendix
%============================================ 
\section{Large sample properties of $Q_k$} 
%============================================ 
We start with noting well-known results about $\hat{F}_k(\cdot)$ and $\hat{R}_k(\cdot).$
As it is shown by \cite{FH1991},
\begin{equation} 
{n}_k^{1/2} \left\{ \frac{\hat{F}_k(\tau) - {F}_k(\tau)}{1-F_k(\tau)}  \right\} = n_k^{-1/2} \sum_{i=1}^{n_k} \int_0^{\tau} \frac{dM_{ki}(u)}{G_{k}(u)} +o_p(1),
\label{eqn:Fdist}
\end{equation}
and this converges weakly to a zero-mean normal distribution,
where
$G_k(t) = \Pr (X_{k} \ge t),$
$M_{ki}(t) = N_{ki}(t) - \int_0^t Y_{ki}(s)dH_{k}(s),$
$N_{ki} (t) = I(X_{ki} \le t, \Delta_{ki} =1), $ and   
$Y_{ki} (t) = I(X_{ki} \ge t),$
$H_k(t)$ is the cumulative hazard function of $T_{k}$. 
Also, from the results shown by \cite{Zhao:2012bp}
\begin{equation} 
{n}_k^{1/2} \left\{\hat{R}_k(\tau) - {R}_k(\tau)  \right\} = - n_k^{-1/2} \sum_{i=1}^{n_k} \int_0^{\tau} 
 \left\{ \int_{u}^{\tau} S_k(t)dt  \right\}
\frac{dM_{ki}(u)}{G_{k}(u)} + o_p(1), 
\label{eqn:Rdist}
\end{equation}
which converges weakly to a zero-mean normal distribution.

Note that we assume that for $\tau_{2}>\tau_{1}\ge 0,$
$F_k(\tau_{2}) > F_k(\tau_{1}) \ge 0$ and 
$R_k(\tau_{2})-R_k(\tau_{1}) > 0.$
Applying the Taylor series expansion to $Q_k$ introduced in (\ref{eqn:Qk1}),

\begin{equation} 
\begin{split}
Q_{k} & =n_{k}^{1/2}\left(\left\{ F_{k}(\tau_{2})-F_{k}(\tau_{1})\right\} ^{-1}\left[\left\{ \hat{F}_{k}(\tau_{2})-\hat{F}_{k}(\tau_{1})\right\} -\left\{ F_{k}(\tau_{2})-F_{k}(\tau_{1})\right\} \right]\right)\\
& -n_{k}^{1/2}\left(\left\{ R_{k}(\tau_{2})-R_{k}(\tau_{1})\right\} ^{-1}\left[\left\{ \hat{R}_{k}(\tau_{2})-\hat{R}_{k}(\tau_{1})\right\} -\left\{ R_{k}(\tau_{2})-R_{k}(\tau_{1})\right\} \right]\right)+o_{p}(1).
\label{eqn:Qk2}
\end{split}
\end{equation}
From the results of (\ref{eqn:Fdist}), 
\begin{equation} 
\begin{split}
&
n_{k}^{1/2}\left[\hat{F}_{k}(\tau_{2})-\hat{F}_{k}(\tau_{1})-\left\{ {F}_{k}(\tau_{2})-{F}_{k}(\tau_{1})\right\} \right]  
\\ 
& = 
n_k^{-1/2} \sum_{i=1}^{n_k} \int_0^{\tau_{2}}
\left\{ 
{S}_k(\tau_{2}) - I(u\le\tau_{1}) {S}_k(\tau_{1}) 
\right\}
\frac{dM_{ki}(u)}{G_{k}(u)} + o_p(1),
\label{eqn:Wf}
\end{split}
\end{equation} 
which converges weakly to a zero-mean normal distribution.
Also, using the results of (\ref{eqn:Rdist}),
\begin{equation} 
\begin{split}
&
n_{k}^{1/2}\left[\hat{R}_{k}(\tau_{2})-\hat{R}_{k}(\tau_{1})-\left\{ {R}_{k}(\tau_{2})-{R}_{k}(\tau_{1})\right\} \right]  
\\ 
& = 
-n_k^{-1/2} \sum_{i=1}^{n_k} \int_0^{\tau_{2}}
\left\{ 
\int_u^{\tau_{2}} {S}_k(t)dt - I(u \le \tau_{1}) \int_u^{\tau_{1}} {S}_k(t)dt 
\right\}
\frac{dM_{ki}(u)}{G_{k}(u)} + o_p(1),
\label{eqn:Wr}
\end{split}
\end{equation} 
which converges weakly to a zero-mean normal distribution.
Incorporating the results of (\ref{eqn:Wf}) and (\ref{eqn:Wr}) into (\ref{eqn:Qk2}),
\begin{equation*}
\begin{split}
Q_k & = n_k^{-1/2} \sum_{i=1}^{n_k} \int_0^{\tau_{2}} 
\left\{ 
\frac 
{{S}_k(\tau_{2}) - I(u \le \tau_{1}) {S}_k(\tau_{1})} 
{{F}_k(\tau_{2})-{F}_k(\tau_{1})} \right. \\
& + \left. \frac
{\int_u^{\tau_{2}} {S}_k(t)dt - I(u \le \tau_{1}) 
 \int_u^{\tau_{1}} {S}_k(t)dt} 
{{R}_k(\tau_{2})-{R}_k(\tau_{1})} 
\right\}
\frac{dM_{ki}(u)}{G_{k}(u)} + o_p(1).
\end{split}
\end{equation*} 
Therefore, by the martingale central limit theorem, it is shown that $Q_k$ converges weakly to a normal distribution with mean 0 and variance
\begin{equation*}
%\begin{split}
V(Q_k) =\int_0^{\tau_{2}}
\left\{ 
\frac 
{{S}_k(\tau_{2}) - I(u \le \tau_{1}) {S}_k(\tau_{1})} 
{{F}_k(\tau_{2})-{F}_k(\tau_{1})}
 + 
\frac {\int_u^{\tau_{2}} {S}_k(t)dt - I(u\le\tau_{1}) \int_u^{\tau_{1}} {S}_k(t)dt} 
{{R}_k(\tau_{2})-{R}_k(\tau_{1})} 
\right\}
^{2}
\frac{dH_{k}(u)}{G_{k}(u)}.
%\end{split}
\end{equation*} 

%%%%%%%%%%%%%%%%
%============================================ 
\section{Large sample properties of $U_k$} 
%============================================ 
To show large sample properties of $U_k$, we rewrite (\ref{eqn:Uk1}) by 
\begin{equation*} 
\begin{split}
U_k & = n_k^{1/2} \left\{\frac{ \hat{F}_k(\tau_{2}) - \hat{F}_k(\tau_{1}) } { \hat{R}_k(\tau_{2}) - \hat{R}_k(\tau_{1})} - \frac{ F_k(\tau_{2}) - F_k(\tau_{1}) }{ \hat{R}_k(\tau_{2}) - \hat{R}_k(\tau_{1}) }\right\} \\
& + n_k^{1/2} \left\{\frac{ F_k(\tau_{2}) - F_k(\tau_{1}) }{ \hat{R}_k(\tau_{2}) - \hat{R}_k(\tau_{1}) } - \frac{ F_k(\tau_{2}) - F_k(\tau_{1}) } {R_k(\tau_{2}) - R_k(\tau_{1})}\right\}.
\end{split}
\end{equation*}
By the application of Taylor series expansion, $U_k$ is denoted by
\begin{equation} 
\begin{split}
& U_k = n_{k}^{1/2}\left[\hat{F}_{k}(\tau_{2})-\hat{F}_{k}(\tau_{1})-\left\{ {F}(\tau_{2})-{F}_{k}(\tau_{1})\right\} \right]\left\{{R}_{k}(\tau_{2})-{R}_{k}(\tau_{1})\right\}^{-1}\\
 & - n_{k}^{1/2}\left\{ {F}_{k}(\tau_{2})-{F}_{k}(\tau_{1})\right\} \left\{ {R}_{k}(\tau_{2})-{R}_{k}(\tau_{1})\right\} ^{-2}\left[\hat{R}_{k}(\tau_{2})-\hat{R}_{k}(\tau_{1})-\left\{ {R}_{k}(\tau_{2})-{R}_{k}(\tau_{1})\right\} \right] \\
 & + o_{p}(1).
\end{split}
\label{eqn:Uk3}
\end{equation} 
Incorporating the results of (\ref{eqn:Wf}) and (\ref{eqn:Wr}) into (\ref{eqn:Uk3}),
\begin{equation*}
\begin{split}
U_k & = n_k^{-1/2} \sum_{i=1}^{n_k} 
\int_0^{\tau_{2}} 
\left[ 
\frac 
{{S}_k(\tau_{2})-I(u\le \tau_{1}){S}_k(\tau_{1})} {{R}_k(\tau_{2}) - {R}_k(\tau_{1})} \right. \\
& + \left. \frac
{ {F}_k(\tau_{2})- {F}_k(\tau_{1}) }
{ \left\{ {R}_k(\tau_{2})- {R}_k(\tau_{1}) \right\}^2}
\left\{ \int_u^{\tau_{2}} {S}_k(t)dt - I(u \le \tau_{1}) \int_u^{\tau_{1}} {S}_k(t)dt \right\} 
\right]
\frac{dM_{ki}(u)}{G_{k}(u)} + o_p(1).
\end{split}
\end{equation*} 
Therefore, by the Martingale central limit theorem, it is shown that $U_k$ converges weakly to a normal distribution with mean 0 and variance
\begin{equation}
\begin{split}
V(U_k) & = 
\int_0^{\tau_{2}} 
\left[ 
\frac 
{{S}_k(\tau_{2})-I(u\le \tau_{1}){S}_k(\tau_{1})} {{R}_k(\tau_{2}) - {R}_k(\tau_{1})} \right. \\
& + \left. \frac
{ {F}_k(\tau_{2})- {F}_k(\tau_{1}) }
{ \left\{ {R}_k(\tau_{2})- {R}_k(\tau_{1}) \right\}^2}
\left\{ \int_u^{\tau_{2}} {S}_k(t)dt - I(u \le \tau_{1}) \int_u^{\tau_{1}} {S}_k(t)dt \right\} 
\right]^2
\frac{dH_{k}(u)}{G_{k}(u)},
\label{eqn:Var_Uk}
\end{split}
\end{equation} 
where $H_k(t)$ is the cumulative hazard function of $T_{k}$. $V(U_k)$ can be estimated by replacing the unknown quantities in (\ref{eqn:Var_Uk}) with their empirical counterparts, as shown below.
\begin{equation*}
\begin{split}
\hat{V}(U_k) & =
\int_0^{\tau_{2}} 
\left[ 
\frac 
{\hat{S}_k(\tau_{2})-I(u\le \tau_{1})\hat{S}_k(\tau_{1})} {\hat{R}_k(\tau_{2}) - \hat{R}_k(\tau_{1})} \right. \\
& + \left. \frac
{ \hat{F}_k(\tau_{2}) - \hat{F}_k(\tau_{1}) }
{ \left\{ \hat{R}_k(\tau_{2})- \hat{R}_k(\tau_{1}) \right\}^2}
\left\{ \int_u^{\tau_{2}} \hat{S}_k(t)dt - I(u \le \tau_{1}) \int_u^{\tau_{1}} \hat{S}_k(t)dt \right\} 
\right]^2
\frac{d\hat{H}_{k}(u)}{\hat{G}_{k}(u)},
\label{eqn:Var_Uk_est}
\end{split}
\end{equation*} 
where $\hat{G}_k(t) =  n^{-1}_k \sum_{i=1}^{n_k} I(X_{ki}\ge t),$ 
and $\hat{H}_k(\cdot)$ is the Nelson-Aalen estimator for $H_k(\cdot)$ for group $k.$

%============================================ 
\section{Connection to the landmark analysis} 
%============================================ 
Let $S_T(t)$ denote the survival function of the event time distribution $T.$ In Section 2, we defined the LT-AH with a time window $(\tau_{1}, \tau_{2})$ as a function of $S_T(t)$ as follows, 
\begin{equation*} 
 {\eta}_k(\tau_{1}, \tau_{2}) = 
 \frac{S_{T}(\tau_{1})-S_{T}(\tau_{2})}{\int_{\tau_{1}}^{\tau_{2}}S_{T}(u)du}.
\end{equation*}
Interestingly, ${\eta}_k(\tau_{1}, \tau_{2})$
can be interpreted as the standard AH over the time window $[0, \tau_2-\tau_1]$ for residual life time $T^*=T-\tau_1$ conditional on $T\ge \tau_1.$ 
Specifically, the standard AH \citep{Uno2023-sm} of $T^*$ with the truncation time $\tau_2 - \tau_1$ is 
\begin{equation*} 
\frac{1-S_{T^*}(\tau_{2}-\tau_{1})}
     {\int_{0}^{\tau_{2}-\tau_{1}}S_{T^*}(u)du} =
\frac{1-S_{T}(\tau_{2})/S_{T}(\tau_{1})}
     {\int_{0}^{\tau_{2}-\tau_{1}}S_{T}(u+\tau_{1})/S_{T}(\tau_{1})du} = 
\frac{S_{T}(\tau_{1})-S_{T}(\tau_{2})}{\int_{\tau_{1}}^{\tau_{2}}S_{T}(u)du},
\end{equation*}
because 
$S_{T^*}(t)=S_{T}(t+\tau_{1})/S_{T}(\tau_{1}).$

The nonparametric estimation and the corresponding inference procedures for 
${\eta}_k(\tau_{1}, \tau_{2})$ introduced in Section 2 (\ref{eqn:LTAH2}) were derived by replacing $S_T(t)$ by its Kaplan-Meier estimator. 
It is equivalent to making inference on the standard AH of $T^*$  
using methods presented by \cite{Uno2023-sm} based on the subgroup of patients whose $X_{ik}>\tau_1.$

%-------------------------------
\section*{Supporting information}
This research was supported by the US NIH grants R01GM152499 (HU) and R01HL089778 (LT), and the McGraw/Patterson Research Fund (HU).

%-------------------------------
\bibliographystyle{biom}
\bibliography{refs}

\end{document}